\documentclass[12 pt]{article}

\usepackage{graphicx}  

\begin{document}

\author{D. Mugnai \\
{\small \it Istituto di Ricerca sulle Onde Elettromagnetiche ``Nello Carrara'' - CNR,} \\ {\small \it Via Panciatichi 64, 50127 Firenze, Italy}}

\title{The tunnel effect in electromagnetic propagation}

\date{}

\maketitle

\begin{abstract}
The tunnel effect is considered here within the framework of 
electromagnetic propagation. The classical problem of a plane gap
of dielectric, surrounded on both sides
by a medium with larger refraction index, is studied in the case in which
an electromagnetic plane wave impinges into the gap with an incidence angle
larger than the critical angle. 
In this condition (total reflection), the gap acts as a classically
forbidden region and behaves like a tunnel.
The field inside the forbidden gap consists of two
evanescent waves, each one having its wavefronts normal to the interface.
In the present paper we study the total field derived as a superposition
of two such evanescent waves, its wavefronts, and the directions of
propagation of both phase and energy.
\end{abstract}

\vspace{.5 cm}

In electromagnetism, an effect analogous
to  the tunnel effect of quantum mechanics occurs when a plane wave,
propagating in a medium with refractive index $n$, impinges into  a
plane-parallel dielectric gap with refractive index $n^\prime$
smaller than $n$, at an incidence angle larger than the critical 
angle $i_0$ defined as $\sin i_0 =n/ n^\prime$.
If the thickness $d$ of the gap is infinite, the field within the gap
consists of a plane evanescent wave -- attenuating in the direction
normal to the interface --  whose phase propagates in
 direction parallel to the 
interface (see Fig. 1a). If $d$ is finite, the boundary conditions on both
interfaces cannot be satisfied by a single evanescent wave and  two
evanescent waves, with the same direction of
propagation of the phase,
but attenuating into opposite directions, are required (see Fig. 1b)
\cite{note}.

\begin{figure}
\begin{center}
\includegraphics[width=.7\textwidth]{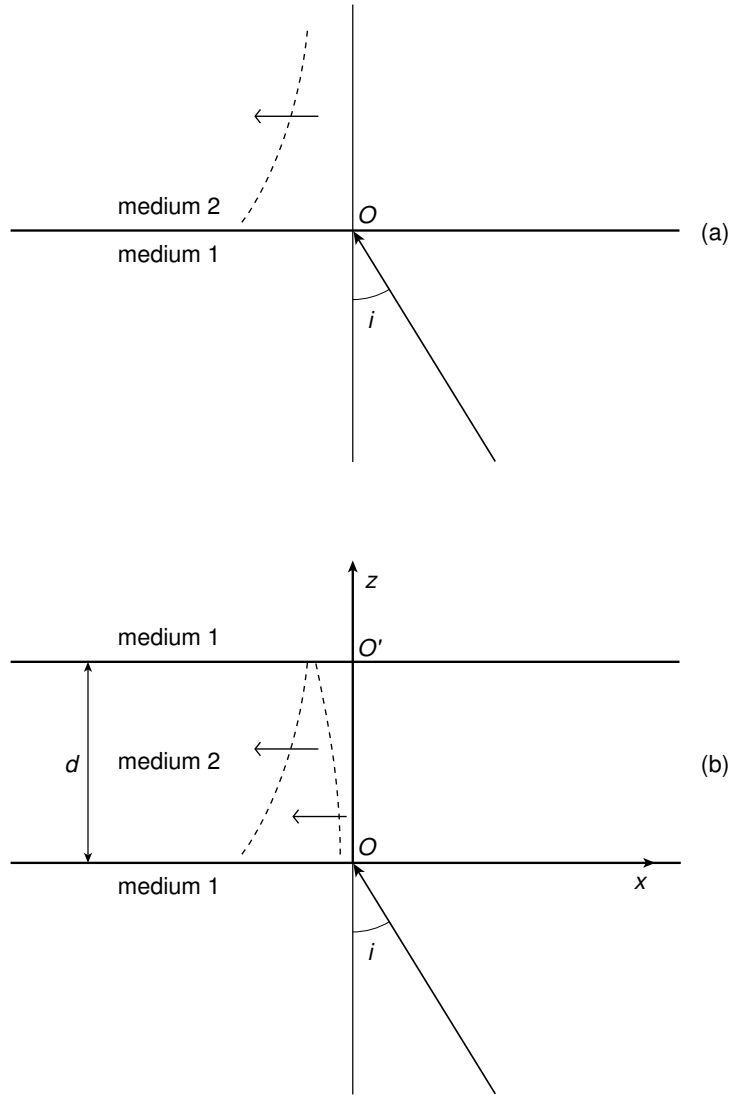}
\end{center}
\caption{\small a) Border surface between medium 1, 
with refractive index $n^\prime$, 
and medium 2, with refractive index $n<n^\prime$. For an incidence angle $i$ 
larger than the critical angle $i_0 =\arcsin (n/n^\prime )$, a single 
evanescent wave originates in medium 2 and propagates parallel 
to the interface.
b) Finite gap thickness of  medium 1, surrounded on both sides by medium 2. 
In this case, two evanescent waves, with the same direction of propagation but
attenuating into opposite directions, originate inside the gap.
The width of the gap $d$ is shown, together with the coordinate system 
adopted in the theoretical analysis.}
\end{figure} 

The properties of the total field inside the forbidden region are due
to the fact that the two evanescent waves to be added have 
real amplitudes with opposite trends of variation in addition to the
fact that the wavefronts, parallel to one another and normal to the
interface,
are not coinciding (on one wavefront the phase of a wave is different
from the phase of the other wave, see Eqs.(\ref{pr})).
Accordingly, the
phase of the total field varies along the wavefronts of the single waves
to be added, and the wavefronts of the total field are not parallel to
those of the component waves.

Let us consider a system of
Cartesian coordinates $x,y,z$ (unit vectors of the axes {\bf i}, {\bf j},
{\bf k}) with origin at $O$ (Fig. 1b), and
an impinging TE wave with the electric field parallel to {\bf j},
with  direction of propagation
{\bf s}$^i =(\alpha${\bf i} $+\gamma${\bf k})
in the plane $xz$.
The $y$-component of the incident electric field can be written as

\begin{equation}   %2
E^i=E_0 \exp [ik_0 n (\alpha x+\gamma z)]\:, 
\end{equation}
where $k_0=\omega /c$ is the free-space wavenumber, $E_0$
(which we assume to be real) denotes
the amplitude of the incident field at the origin $O$ and

\begin{eqnarray*}
\gamma =\sqrt{1-\alpha^2} \:. 
\end{eqnarray*}
Inside the gap the total field is the superposition of two TE 
evanescent waves, whose electric field (still parallel to {\bf j}) 
can be written as 

\begin{eqnarray} %3
E^+ &=& pE_0 \exp [k_0(in\alpha x- \Gamma z)]\exp (-i\omega t)\:, \nonumber \\
E^- &=& rE_0 \exp [k_0(in\alpha x+ \Gamma z)]\exp (-i\omega t) \:,
\end{eqnarray}
where

\begin{eqnarray}
\Gamma =\sqrt{n^2\alpha^2 -1 } \nonumber
\end{eqnarray}
is a real quantity if, as assumed, the incidence angle is larger 
than the critical angle ($\alpha > 1/n$). 

The complex coefficients 
$p$ and $r$ can be deduced from the boundary conditions on the 
interfaces at $z=0$ and $z=d$\cite{atti} 

\begin{eqnarray}
p= \frac{e_2(\Gamma -in\gamma)}{2\Gamma}\, \tau =|p| \exp (i\varphi_p)
\nonumber \\
r= \frac{e_1(\Gamma +in\gamma)}{2\Gamma}\, \tau =|r| \exp (i\varphi_r)\:,
\end{eqnarray}
where

\begin{eqnarray} %4
e_1 &=& \exp (-k_0\Gamma d)\:,\:\:\:\:\:
e_2= \frac{1}{e_1}=\exp (k_0\Gamma d) \nonumber \\
\tau &=& |\tau | \exp (i\varphi_\tau )=\frac{4in\gamma\Gamma}{e_1
(\Gamma +in\gamma )^2 - e_2 (\Gamma -in\gamma )^2}
\end{eqnarray}
and

\begin{eqnarray} %5
|p| &=& \frac{e_2}{2\Gamma}\,\Delta |\tau |   \:,\:\:\:\:\:
|r| =\frac{e_1}{2\Gamma}\,\Delta |\tau |   \nonumber \\
\varphi_p &=& \varphi_\tau -\Phi \:, \:\:\:\: \varphi_r = 
\varphi_\tau +\Phi \nonumber \\
\Delta &=& \sqrt{\Gamma^2+n^2\gamma^2} = \sqrt{n^2-1}  \nonumber \\
\Phi &=& \arctan \left(\frac{n\gamma}{\Gamma}\right) \:.
\label{pr}
\end{eqnarray}
From Eq. (\ref{pr}) it turns out that $\varphi_p$ and
$\varphi_r$ differ by $2\Phi$.
All the above quantities are independent of the coordinates,
and $\tau$ represents the amplitude transmission coefficient
of the gap\cite{atti}.
The total electric field inside the gap can therefore be written as
(apart from the time dependence $\exp (-i\omega t)$)

\begin{eqnarray} %6
E^g= E^++E^- &=& \frac{E_0\Delta |\tau |}{2\Gamma}\, \exp (ik_0n\alpha x)
\times \nonumber \\
 &\times  & \left[ \frac{}{}\exp [k_0\Gamma (d-z)]\exp (i\varphi_p)+
\exp [-k_0\Gamma (d-z)]\exp (i\varphi_r)\frac{}{}\right] 
\end{eqnarray}
and, by taking into account Eq.(\ref{pr}), we have

\begin{eqnarray} %7
 E^g &=& \frac{E_0\Delta |\tau |}{2\Gamma}\, \exp (ik_0n\alpha x
 +i\varphi_\tau )
 \times \nonumber \\
&\times &\left[ \frac{}{} \exp [k_0\Gamma (d-z)]\exp (-i\Phi)+
\exp [-k_0\Gamma (d-z)]\exp (i\Phi) \frac{}{} \right]\:.
\label{eg}
\end{eqnarray} 
From Eq.(\ref{eg}) it turns out that the amplitude of $E^g$
depends only on $z$ (the equi-amplitude surfaces are the
planes $z=$ constant), and  decreases from the value
$ E_0 \Delta |\tau |\sqrt{\cosh^2(k_0\Gamma d)-\sin^2\Phi}/\Gamma$,
at $z=0$, to the value
$ E_0 |\tau |$, at $z=d$.
The spatial dependence of the amplitude implies that geometrical
optics is inadequate for describing electromagnetic propagation within
a tunneling region\cite{diffra}.

As to the phase, from Eq. (\ref{eg}) it follows that the equation of
wavefronts (equi-phase surfaces) is given by

\begin{equation}  %8
\varphi (x,z)= k_0 n \alpha x +\Psi (z)+ \varphi_\tau = \varphi_0 \:,
\label{fi}
\end{equation}
where $\varphi_0$ is a constant and

\begin{eqnarray} %9
\tan [\Psi (z)]&=& -\tanh [k_0\Gamma (d-z)]\tan (\Phi )=
- \, \frac{n\gamma}{\Gamma} \tanh [k_0\Gamma (d-z)] \nonumber \\
\varphi_\tau &=& \arctan \left[ \frac{n^2\gamma^2-\Gamma^2}{2n\gamma\Gamma}
\, \tanh (k_0\Gamma d)\right] \:.
\end{eqnarray}
Looking at Eq. (\ref{fi}), we can see that different wavefronts,
corresponding
to  different values of the phase, are simply shifted in the $x$-direction
with respect to one another.
Figure 2 shows the wavefronts $\varphi_0 =0,\pi ,2\pi$
for parameter values referring to an experiment dealing with
frustrated total reflection in the range of microwaves\cite{prism}.

\begin{figure}
\begin{center}
\includegraphics[width=.65\textwidth]{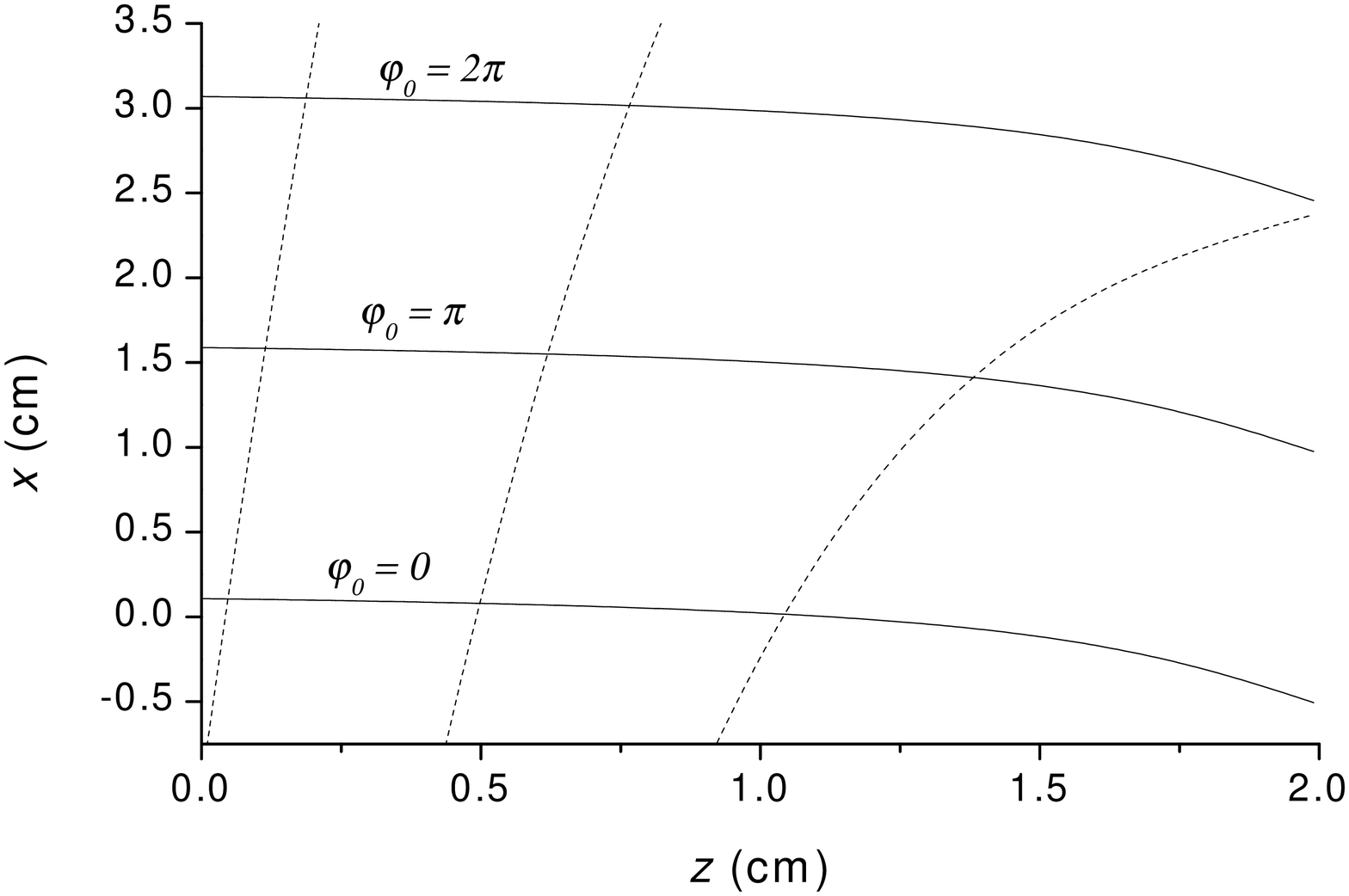}
\end{center}
\caption{\small Wavefronts (solid lines) as derived from Eq. (\ref{fi}),
for $\varphi_0 =0,\:\pi ,\:2\pi$, and
rays (dashed lines) as derived from Eq. (\ref{x}), for
three arbitrary values of the constant $X_0$.
The parameter values  refer to
an experimental situation\cite{prism} in which 
medium 1  consists of paraffin ($n^\prime =1.49$) and medium 2 of  air ($n=1$). Other parameter values are: $\alpha = 0.68,\: 
\nu =10 \: {\rm GHz}\: (\omega = 2\pi\nu ),\: c= 30 \:{\rm cm/ns}$.}
\end{figure}

We can now derive both the wavelength and
the equation of the rays. To this end, we have to evaluate
grad$[\varphi (x,z)]$
and we have

\begin{equation}    %10
\lambda = \frac{2\pi}{|{\rm grad}[\varphi (x,z)]|}
\label{lam}
\end{equation}
and, for the ray equation $x=X(z)$ defined as the lines of flux of
grad$\varphi$,

\begin{equation} %11
\frac{dX}{dz}= \frac{(\partial\varphi /\partial x)}
{(\partial \varphi /\partial z)} \:.
\end{equation} 
From Eq. (\ref{fi}) we have

\begin{equation} %12
\frac{\partial\varphi}{\partial x}= k_0 n\alpha
\end{equation}

\begin{eqnarray}  %13
\frac{\partial\varphi}{\partial z}&=& \frac{k_0\Gamma \sin\Phi\,\cos\Phi}
{\cos^2\Phi\cosh^2[k_0\Gamma (d-z)]+\sin^2\Phi\sinh^2 [k_0\Gamma (d-z)]}
\nonumber \\
&=& \frac{k_0\Gamma\sin\Phi\cos\Phi}{\cosh^2[k_0\Gamma (d-z)]-\sin^2\Phi}
\end{eqnarray}
hence

\begin{eqnarray} %14
|{\rm grad}\,\varphi | &=& k_0\sqrt{n^2\alpha^2+
\frac{\Gamma^2\sin^2\Phi\cos^2\Phi}
{\left[\cosh^2[k_0\Gamma (d-z)]-\sin^2\Phi \right]^2}} 
\label{grad}
\end{eqnarray}
and

\begin{eqnarray}   %15
\frac{dX}{dz} &=& \frac{n\alpha}{\Gamma \sin\Phi\,\cos \Phi}
\left[  \cosh^2[k_0\Gamma (d-z)]-\sin^2\Phi \right] \nonumber \\
&=& \frac{\alpha \Delta^2}{\gamma\Gamma^2} \left[\cosh^2[k_0\Gamma (d-z)]-
\frac{n^2\gamma^2}{\Delta^2}\right]  \:.
\label{dx}
\end{eqnarray}
From Eq. (\ref{grad}) it follows that $|{\rm grad}\,\varphi |>k_0n\alpha$ and,  
from Eq. (\ref{lam}),

\begin{equation} %16
\lambda < \frac{\lambda_0}{n\alpha}< \lambda_0, \:\:\:\:\:\: (n\alpha >1)  
\label{lam0}
\end{equation}
where $\lambda_0 =2\pi /k_0$ is the free-space wavelength
(medium 2 in Fig. 1).
We can conclude, therefore, that the total
field inside the tunneling region is slow, that is
the phase velocity along the rays is slower than the light velocity $c$.
By integrating Eq. ({\ref{dx}), we obtain the ray equation

\begin{equation} %17
X(z)= \frac{\alpha\Delta^2}{\gamma\Gamma^2}
\left[ z\left( \frac{1}{2}- \frac{n^2\gamma^2}{\Delta^2} \right)
-\frac{1}{4k_0\Gamma} \sinh [2k_0\Gamma (d-z)]  \right] + X_0\:, 
\label{x}
\end{equation} 
where $X_0$ is a constant.
From Eq. (\ref{x}), we can see that (as was to be expected due to the
symmetries of the problem) the rays are shifted in the $x$-direction,
and that, since the amplitude of the total field is not constant with
respect to $z$, they  are not straight lines (see Fig. 2).

By denoting the angle between a ray
and the $z$-axis with $\chi$
($\tan\chi = dX/dz$ (see Eq. (\ref{dx})), we have, at $z=0$,

\begin{eqnarray*}
\sin\chi = \frac{\Delta^2\cosh^2(k_0\Gamma d)-n^2\gamma^2}
{\sqrt{ \left[\Delta^2\cosh^2(k_0\Gamma d)-n^2\gamma^2\right]^2
+\left( \frac{\gamma\Gamma^2}{\alpha}\right)^2 }}
\end{eqnarray*}
while, at $z=d$,

\begin{eqnarray*}
\sin \chi =\alpha
\end{eqnarray*}
Since $\chi$ is the refraction angle at $z=0$ and the
incidence angle into the second interface at $z=d$, the unexpected
conclusion is thus that the refraction low seems not to be
valid for the rays inside the gap.

The Poynting vector inside the gap has  a component
normal to the interfaces, contrarily to what happens for the field
on the right of the single interface of Fig. 1.
In order to evaluate the flux lines of the Poynting vector and
their relationship with the flux lines of the phase,
let us consider the vector {\bf S} describing the energy
propogation\cite{onde}

\begin{equation}  %18
{\rm \bf S}= \frac{1}{2} {\rm Re} ({\rm\bf E}\wedge {\rm\bf H}^\star )\:,
\end{equation}
where the asterix indicates a complex conjugate. We easily obtain

\begin{eqnarray}
S_x &=& \frac{1}{2 Z_0} n\alpha \,\left(\frac{E_0\Delta |\tau |}{\Gamma}
\right)^2 \left[
\cosh^2 [k_0\Gamma (d-z)]-\sin^2\Phi  \right] \nonumber \\
S_z &=& \frac{1}{2 Z_0} n\gamma \, \left( E_0 |\tau |\right)^2 \:,
\end{eqnarray}
where $Z_0$ is the free space impedance. 
As expected, the $z$-component of the Poynting vector does not depend on $z$.
The flux lines of the Poynting vector can therefore be written as

\begin{equation}
\frac{dx}{dz} = \frac{S_x}{S_z} = \frac{\alpha\Delta^2}{\gamma\Gamma^2}
\left[  \cosh^2[k_0\Gamma (d-z)] -\frac{n^2\gamma^2}{\Delta^2} \right]
\end{equation}
and, by comparing this with Eq. (\ref{dx}), we see that the flux lines of the
energy coincide with the flux lines of the phase.

By means of Eq. (\ref{fi}),  we are able to evaluate the phase
difference $\Delta\varphi$ between two opposite points, $O$ and $O^\prime$,
along the
$z$-direction inside the gap (see Fig. 1b): we have

\begin{eqnarray*}
\Delta\varphi = \varphi (0,d)-\varphi (0,0) = \arctan\left[
\frac{n\gamma}{\Gamma} \tanh (k_0\Gamma d) \right] \:.
\end{eqnarray*}
By including in the total phase also the temporal factor
$\exp (-i\omega t)$ disregarded until now, the phase delay
in going from $O$ (at $t=0$) to $O^\prime$ is

\begin{equation}  %21
t_\varphi= \Delta\varphi /\omega.
\label{tfi}
\end{equation}
In Fig. 3, we show $t_\varphi$ as a function of the gap width,
together with the time $t_l=d/c$. 
If we suppose to perform an experiment (like the one reported in
Ref. \cite{prism})
with a monocromatic wave and we put two probes at $O$ and $O^\prime$,
we would actually measure a phase delay as given by
Eq. (\ref{tfi}) which is,
without doubt, an observable\cite{note1}. 
To clarify the physical meaning of
this delay we have to complete the analysis by considering not a plane wave impinging the gap, but a narrow beam or a wave packet\cite{bos,ste}. 
However, this improvement exceeds the purpose of the present work and will be presented elsewhere.   

\begin{figure}
\begin{center}
\includegraphics[width=.7\textwidth]{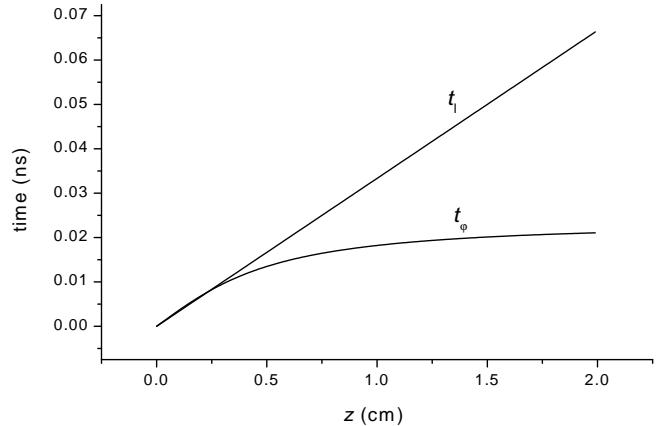}
\end{center}
\caption{{\small Phase delay time $t_\varphi$ along the $z$-direction 
as derived from Eq. (\ref{tfi}), together with the time $t_l=z/c$, 
as a function of $z$. Parameter values are the same as those in Fig. 2.}}
\end{figure} 

Finally, we wish to note that $t_\varphi$ becomes independent on $d$,
for large $d$ (see Eq. (\ref{tfi})).
A behaviour of this kind was also obtained within
the framework
of a quantum-mechanical theoretical model, due to Hartman\cite{har}, for
a particle tunneling through a rectangular potential barrier.
Also in that case, the ``traversal time'' under barrier
tends to be  constant for large barriers, and the superluminal effect
so obtained is known as "Hartman effect".
It is not easy to understand the nature of that time but 
its behaviour, for large barrier, very similar to the one as 
derived here (Eq. (\ref{tfi})), 
could characterised it as a phase-delay\cite{prism,diffra1}.

\vspace{1 cm}
\noindent
{\it Acknowledgments}

Thanks are due to L. Ronchi Abbozzo and A. Ranfagni for useful 
discussions and suggestions.


\begin{thebibliography}{50}
\bibitem{note1}
From the boundary conditions, that is continuity conditions for the
tangential component of the electric fields across the interfaces,
at $z=d$ (in $O^\prime$)
we have that the transmitted field is exactly equal to the internal field.
\end{thebibliography}
\end{document}